\documentclass[aps,prb,twocolumn,longbibliography]{revtex4-2}

\usepackage{graphicx}
\usepackage[dvipsnames]{xcolor}
\usepackage[colorlinks=true, allcolors=RoyalBlue]{hyperref}
\usepackage{amsmath}
\usepackage{amssymb}
\usepackage{soul}
\usepackage{hyperref}
\usepackage[english]{babel}

\begin{document}

\title{Doped Mott phase and charge correlations in monolayer 1T-NbSe$_2$}
\author{Xin Huang}
\affiliation{Department of Applied Physics, Aalto University, FI-00076 Aalto, Finland}

\author{Jose L. Lado}
\affiliation{Department of Applied Physics, Aalto University, FI-00076 Aalto, Finland}

\author{Jani Sainio}
\affiliation{Department of Applied Physics, Aalto University, FI-00076 Aalto, Finland}

\author{Peter Liljeroth}
\affiliation{Department of Applied Physics, Aalto University, FI-00076 Aalto, Finland}

\author{Somesh Chandra Ganguli}
\email{somesh.ganguli@aalto.fi}
\affiliation{Department of Applied Physics, Aalto University, FI-00076 Aalto, Finland}

\begin{abstract}
The doped Hubbard model is one of the paradigmatic platforms to engineer exotic quantum
many-body states, including charge-ordered states, strange metals and unconventional superconductors. 
While undoped and doped correlated phases have been experimentally realized in a variety of
twisted van der Waals materials, experiments in monolayer materials, and in particular 1T
transition metal dichalcogenides, have solely reached the conventional
insulating undoped regime. 
Correlated phases in monolayer two-dimensional
materials have much higher associated energy scales than
their twisted counterparts, making doped correlated monolayers
an attractive platform for high temperature correlated quantum matter.
Here, we demonstrate the realization of a doped Mott phase in a van der Waals dichalcogenide 1T-NbSe$_2$ monolayer. 
The system is electron doped due to electron transfer from a monolayer van der Waals substrate via proximity, 
leading to a correlated triangular lattice with both half-filled and
fully-filled sites. We analyze the distribution of the half-filled and filled sites and show the arrangement is unlikely to be controlled by disorder alone, and we show that the presence of competing non-local many-body correlations would account for the charge correlations
found experimentally. Our results establish 1T-NbSe$_2$ as a potential monolayer platform to explore correlated doped Mott physics in a frustrated lattice.

\end{abstract}

\maketitle

Mott insulators and the Hubbard model \cite{1963,Arovas2022} are one of the paradigmatic platforms
to engineer exotic quantum many-body states.
Doped Mott physics is known to lead to a variety of unconventional
strongly correlated phenomena \cite{Qin2022,RevModPhys.78.17}, 
and represent a key physical regime leading to high-$T_C$ superconductivity. 
Conventional doped Mott systems such as cuprates \cite{RevModPhys.72.969}
feature square lattices and its phenomena
may be impacted by disorder \cite{PhysRevLett.94.056404,PhysRevLett.117.146601},
while more exotic phenomena can potentially
appear in doped Mott systems lacking disorder and
featuring geometrically frustrated lattices \cite{Watanabe2005,PhysRevLett.113.246405,PhysRevB.104.L121118,PhysRevB.96.205130}.
Van der Waals materials \cite{Novoselov2016},
including graphene and dichalcogenides,
have risen as ideal materials systems for filling this gap.
Electrostatic gating in these materials
allows doping correlated systems without disorder effects \cite{Novoselov2004}, 
and their effective triangular
lattices provide a complementary platform to explore the impact of geometric frustration \cite{PhysRevLett.100.136402,PhysRevLett.105.267204}. 
In particular, twisted van der Waals heterostructures have become the preferred
system to emulate correlated states, ranging from conventional
correlated insulators \cite{Cao2018mott,Cao2018}, charge ordered states \cite{RubioVerd2021,Jiang2019}
unconventional superconducting states \cite{Cao2021,Kim2022}
and fractional topological states\cite{Cai2023,Park_2023,Zeng_2023}. While the capability of doping
van der Waals systems
extends to monolayers, tunable correlated physics in monolayers
have remained much less explored.

Monolayer dichalcogenides \cite{Manzeli2017} can be used to realize different correlated states.
Dichalcogenide materials have two typical different crystal structures (1H and 1T phases), with metallic dichalcogenides featuring charge density wave (CDW) orders in both phases. While monolayer
1H phases remain metallic and superconducting at lower
temperatures \cite{Ugeda2015,Vano2023,PhysRevB.99.161119,Khestanova2018}, 1T phases develop a
correlated insulating state \cite{PhysRevB.73.073106,Cho2016,Chen2020,Nakata2021} due to the superstructure
generated by the charge density wave reconstruction \cite{Liu2021}. 
Interestingly, the complex electronic structure
of such a system can effectively result in a single orbital model in triangular lattice supercell \cite{PhysRevB.105.L081106,Chen2022},
leading to a strongly correlated state due to the reduced bandwidth. 1T
dichalcogenides are well-known Mott insulators, and due to the geometric frustration,
they have been proposed as quantum spin liquids \cite{Law2017,Ruan2021}. Their van der Waals nature
turns them into ideal candidates to use as magnetic building block in 
van der Waals heterostructures \cite{Vano2021,Wan2023,Ayani2023}.
Proximity effects can enable doping via charge transfer \cite{2023arXiv230214072C},
turning them into an attractive system to explore doped Mott physics in a triangular lattice \cite{PhysRevB.81.224505,PhysRevB.105.205110,PhysRevB.88.041103,PhysRevB.90.165135,PhysRevB.100.060506}. 
However, doped Mott phases featuring correlated order have remained unexplored experimentally
in monolayer materials.

Here, we  demonstrate the realization of a doped Mott phase
in a van der Waals dichalcogenide monolayer. We show that,
due to the charge reconstruction, an emergent correlated triangular lattice appears. The combination of local
many-body correlations and charge transfer leads to a triangular lattice featuring both half-filled and
fully-filled sites, realizing a deep doped Mott phase. We analyze the statistical distribution
of half-filled and filled sites, showing that the arrangement is unlikely to be controlled by disorder alone
but signaling the presence of underlying many-body non-local correlations. Our results establish 1T-NbSe$_2$
as a potential monolayer platform to explore correlated doped Mott physics in a frustrated lattice.

\begin{figure*}[t!]
    \centering
    \includegraphics[width=0.8\textwidth]{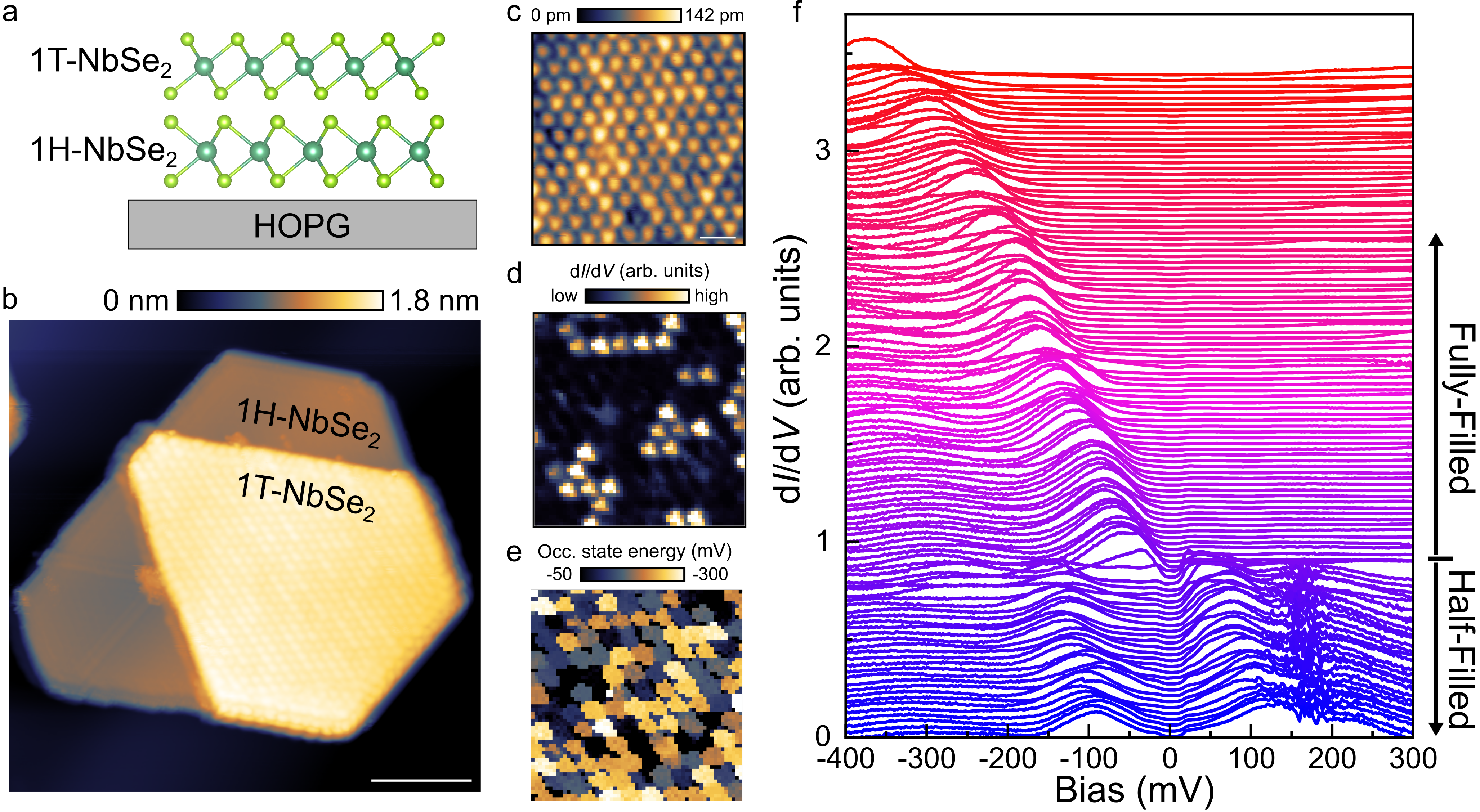}
    \caption{Electronic structure of 1T-NbSe$_2$ on 1H-NbSe$_2$. (a) Schematic side view of the vertical heterostructure. (b) STM topography of the 1T on 1H NbSe$_2$ (bias voltage: -520 mV, set point: 2 pA. Scale bar 10 nm). (c) 1T-NbSe$_2$ topography having star of David CDW (bias voltage: -400 mV, set point: 3 pA. Scale bar 2 nm). (d, e) d$I$/d$V$ map at 74 mV (d) and map of highest occupied state energies (e) over the same area. (f) All the d$I$/d$V$ spectra taken at the centre of the star of David arranged monotonically with respect to the location of upper Hubbard band. The spectra are vertically offset for clarity.}
    \label{fig:fig1}
\end{figure*}

Fig.~\ref{fig:fig1}a shows a schematic of our experimental system where we have grown heterostructures of 1T- and 1H-crystal polymorphs of monolayer NbSe$_2$ via molecular-beam epitaxy (MBE) on highly-oriented pyrolytic graphite (HOPG), see Supplementary Information (SI) for details of the sample growth. The growth yields a statistical distribution of different heterostuctures and we focus here on the 1T/1H-heterobilayers. The samples were investigated by low-temperature scanning tunneling microscopy (STM) at $T=350$ mK and Fig.~\ref{fig:fig1}b shows an STM image of a typical heterostructure. We can identify the 1H and 1T-phases through atomic scale imaging \cite{Ugeda2015,Ganguli2022,Liu2021}. In a zoomed-in image (Fig.~\ref{fig:fig1}c), the lattice set-up by the well-known star-of-David (SoD) CDW reconstruction in 1T-NbSe$_2$ is clearly visible. Similarly to 1T-TaSe$_2$ and 1T-TaS$_2$, each CDW unit cell hosts a single state close to the Fermi level and electron-electron interactions break these into lower and upper Hubbard bands (LHB and UHB). In contrast to well-studied 1T-TaS$_2$ and 1T-TaSe$_2$, the LHB in 1T-NbSe$_2$ has been reported to be located below the valence band, making the material a charge transfer insulator \cite{Liu2021}. 

The doping of the CDW lattice is controlled by the substrate. Monolayer 1T-NbSe$_2$ grown on HOPG is undoped and each CDW unit cell hosts a single unpaired electron (see SI). In heterostructures with 1H-NbSe$_2$ the situation is drastically altered, and only about 30\% of the CDW unit cells remain singly occupied. This is visualized in Fig.~\ref{fig:fig1}f, which shows d$I$/d$V$ spectra recorded at the centers of the SoD unit cells over the area shown in Fig.~\ref{fig:fig1}c. The spectra are ordered in energy, their spatial pattern is complex and disordered (Fig.~\ref{fig:fig1}d,e, see below for details). On the lower end of the plot (indicated half-filled), there are roughly symmetric peaks around the Fermi level corresponding to the LHB and UHB. While monolayer 1T-NbSe$_2$ directly grown on HOPG is a charge transfer insulator, the 1T-monolayer on 1H-NbSe$_2$ shows the typical behaviour of a Mott insulator: as the doping is increased (moving up in Fig.~\ref{fig:fig1}f), the UHB moves towards the Fermi level until it crosses it, the LHB becomes doubly filled and the spectral peak corresponding to the UHB disappears.

The complex spatial distribution of the half- and fully-filled cells can be probed by STS experiments as shown in Fig.~\ref{fig:fig1}d and e. The half-filled unit cells are clearly visible in the d$I$/d$V$ map shown in Fig.~\ref{fig:fig1}d. The energy of the highest occupied band extracted from grid spectroscopy data over the same area is shown in Fig.~\ref{fig:fig1}e. Surprisingly, there is obvious clustering of the half-filled sites, which we analyze more quantitatively below. This is unexpected as in the absence of disorder and interactions, the half-filled sites should be randomly distributed. On the other hand, charge interaction between neighboring SoD unit cells can give rise to ordered patterns of the half-filled and fully-filled unit cells.

\begin{figure*}
    \centering
    \includegraphics[width=.9\textwidth]{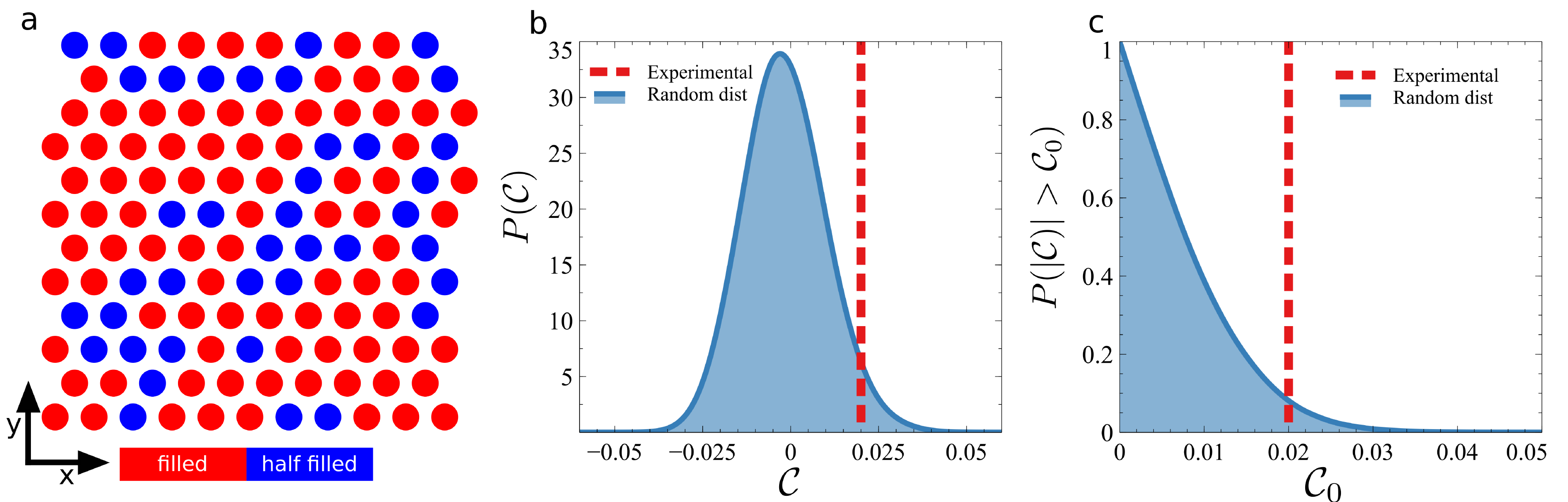}
    \caption{Analysis of the charge distribution in doped 1T-NbSe$_2$. (a) Experimental distribution of half-filled and fully-filled sites. (b) Probability distribution $P(\mathcal{C})$ of the first neighbor correlator $\mathcal{C}$ in the case the fillings of the different sites are purely random, compared with the correlator obtained experimentally. (c) Probability distribution $P(|\mathcal{C}|>\mathcal{C}_0)$ that a purely random arrangement leads to a correlator bigger than a threshold $\mathcal{C}_0$, compared with the experimentally measured value.}
    \label{fig:fig2}
\end{figure*}

The distribution of the half- and fully-filled unit cells can be quantified by calculating spatial correlators of the filled and half-filled cells. The local nearest-neighbor
spatial correlation can be defined as
$\mathcal{C}=\langle n_i n_j \rangle_{ij\in NN} - \langle n_i \rangle \langle n_j \rangle$,
where $n_i=1,2$ for a half-filled/fully-filled site $i$,
$ij\in NN$ denotes nearest neighbor pairs in the SoD triangular
superlattice and $\langle \rangle$ denotes the statistical average. 
For an infinitely large system with a random distribution of fillings, the previous correlators would
be identically $0$. The presence of a finite value of the correlator signals that interactions lead to a preference in the
fillings between neighboring sites.
We have converted the data in Fig.~\ref{fig:fig1} into a binary filled/half-filled format (Fig.~\ref{fig:fig2}a). This is then used to calculate the first neighbor correlator yielding a value of $C=0.02$ indicating significant positive correlations in finding half-filled unit cells next to each other. 
For a finite system, statistical fluctuations can lead to different non-zero values of the correlators even if the distribution
of fillings were purely random. 
Such a purely random doping distribution would arise if the distributions of fillings
is dominated by the effect of impurities or defects in the underlying 1H-NbSe$_2$ substrate.
To quantify the impact of finite size effects, we compare the value of the correlator to the results from a purely random distribution in the finite size geometry of our experiment (Fig.~\ref{fig:fig2}b and c), indicating that the probability of
finding our experimental correlator or bigger
in a random system is less than $10\%$. 
The previous phenomenology shows that our system presents correlations that are unlikely to stem
from randomness alone, and signals the presence of a mechanism leading to finite
interactions between the filling of the sites. In the following we present a mechanism that would account for these
finite correlations. 

The strongly interacting model realized in 1T-NbSe$_2$ is expected not only to feature local electronic repulsion,
but also finite interactions between SoD sites may be present. In general, the interacting model
can be written as extended Hubbard model in the triangular lattice \cite{Watanabe2005,PhysRevLett.113.246405} of the form
\begin{equation}
\begin{aligned}
\mathcal{H} = 
\sum_{i,j,s} t_{i,j} \hat c^\dagger_{i,s} \hat c_{j,s}
+ U 
\sum_{i} 
\left ( \hat n_{i,\uparrow} - \frac{1}{2} \right )
\left ( \hat n_{i,\downarrow} - \frac{1}{2} \right ) \\
+
\sum_{i,s} \mu_i \hat c^\dagger_{i,s} \hat c_{i,s}
+
\sum_{i,j,s,s'} V_{i,j} \hat n_{i,s} \hat n_{j,s'}
\label{eq:UV}
\end{aligned}
\end{equation}
where $V_{i,j}$ parametrizes the nearest neighbors interactions and $\hat n_{i,s} = \hat c^\dagger_{i,s} \hat c_{i,s} $ is the number operator
for spin $s$ in site $i$. 
The hoppings $t_{i,j}$ account for the finite dispersion of the SoD band, $U$ accounts for the local electrostatic repulsion.
Local disorder stemming from the stacking or defects is included in $\mu_i$, where $\mu_i \in [-W,W]$ are random values
with $W$ parametrizing the strength of disorder.
The extended electrostatic interactions $V_{i,j}$ are instrumental to give rise to nearest neighbor correlations in the fillings
compatible with the experimental findings.
The previous model can be solved using a mean field decoupling using a self-consistent procedure \cite{pyqula}, which
allows us to simultaneously capture the impact of local interactions, non-local interactions, charge transfer, kinetic energy
and disorder. Calculations are performed at the electrostatic filling found experimentally,
and we take $U=2.2\beta$, with $\beta$ the bandwidth of the SoD flat band leading to a deep Mott state,
and $V_2 = 0.2U$. 
During the minimization domains and topological defects may appear,
which are directly accounted by the self-consistent solution of the interacting problem.

\begin{figure*}[t!]
    \centering
   \includegraphics[width=\textwidth]{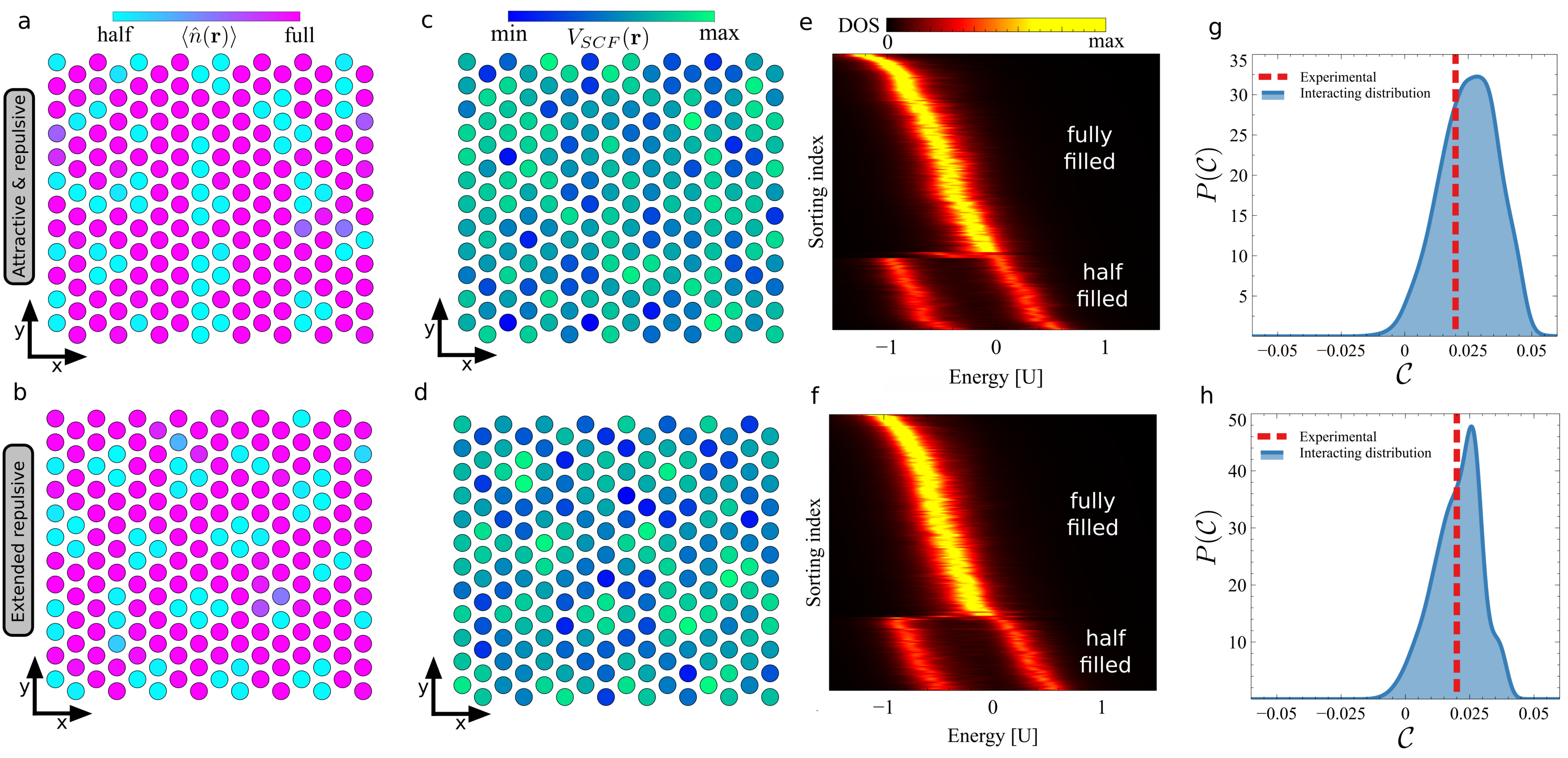}
    \caption{Self-consistent solution of an interacting
    long-range doped Hubbard model with disorder.
    Panels (a,c,e,g) correspond to a model with
    competing attractive and repulsive interactions
    $V_1 = -0.5 V_2$, $V_2>0$ and (b,d,f,h) to a model
    with only repulsive interactions with $V_1=0.3 V_2$, $V_3=V_2$.
    Both models give rise
    to a charge arrangement with positive first neighbor correlator
    $C>0$ as found experimentally (a,b),
    featuring a combination of disorder and many-body interactions giving rise
    to a fluctuating local potential (c,d).
    The density of states is strongly site dependent, giving rise
    to half-filled sites with a Mott gap and fully filled sites (e,f)
    Panels (g,h) show the statistical distribution of the
    correlators for different disorder
    realizations, featuring an overall positive correlation
    both for attractive and repulsive models.
    We took disorder $W=5 V_2$ in (a,c,e,g) and $W=3 V_2$ (b,d,f,h).
   } 
    \label{fig:theory}
\end{figure*}

Let us first consider the minimal case where only local interaction $U$ and nearest neighbor repulsion $V_1$ are present.
In this situation, charges are expected to repel each other
and give rise to a negative first neighbor correlation $\mathcal{C}$. For more complex interactions,
positive correlations can be found. We show in Fig.~\ref{fig:theory} two minimal models consistent
with the experiment, a model with competing
attractive and repulsive interaction $V_1-V_2$, where the nearest neighbor attraction $V_1$
can stem from SoD polarons \cite{Koschorreck2012,PhysRevLett.102.230402,PhysRevLett.103.170402,PhysRevX.13.011029} (Figs.~\ref{fig:theory}a,c,e,g),
and a model with purely repulsive $V_1-V_2-V_3$ interactions \cite{PhysRevB.98.125308,Chen2022} (Fig.~\ref{fig:theory}b,d,f,h). 
For both models, the self-consistent charge arrangement shows distributions
comparable to the experimental ones (Fig.~\ref{fig:theory}a,b) featuring small clusters.
The combination of the local disorder
and the mean-field interactions gives rise to a locally modulated self-consistent
electrostatic potential (Fig.~\ref{fig:theory}c,d),
yielding a spatially dependent density of states in the system (Fig.~\ref{fig:theory}e,f). The
density of states features both half-filled and fully-filled sites (Fig.~\ref{fig:theory}) consistent
with experimental results.

Due to the inherent frustration of the lattice and interactions the ground state may be multiply degenerate, and most importantly depends on the disorder configuration considered. To account for this variability and show the robustness
of this phenomenology we have performed a statistical analysis of the nearest neighbor correlator
by finding the ground state of the interacting model over $10^5$ different disorder configurations. 
The calculations of the nearest neighbor correlator
can be directly performed with a downfolded model that integrates out charge fluctuations.
In the absence of charge fluctuations the long-range interacting model
can be mapped to
a gas-lattice model of the form
$
H = 
\sum V_{ij} \tilde n_i \tilde n_j
$
where $\tilde n_i = 1,2$ are the classical occupations of the sites,
whose lowest energy configuration can be obtained by a Markov chain Monte Carlo minimization.
This classical limit allows us to effectively sample
over many disordered configurations, demonstrating
that generic disorder configurations feature a positive correlation
driven by interactions as shown in Fig.~\ref{fig:theory}g,h.
A more detailed analysis of the model can be found in the SI.

\begin{figure}
    \centering
    \includegraphics[width=1\columnwidth]{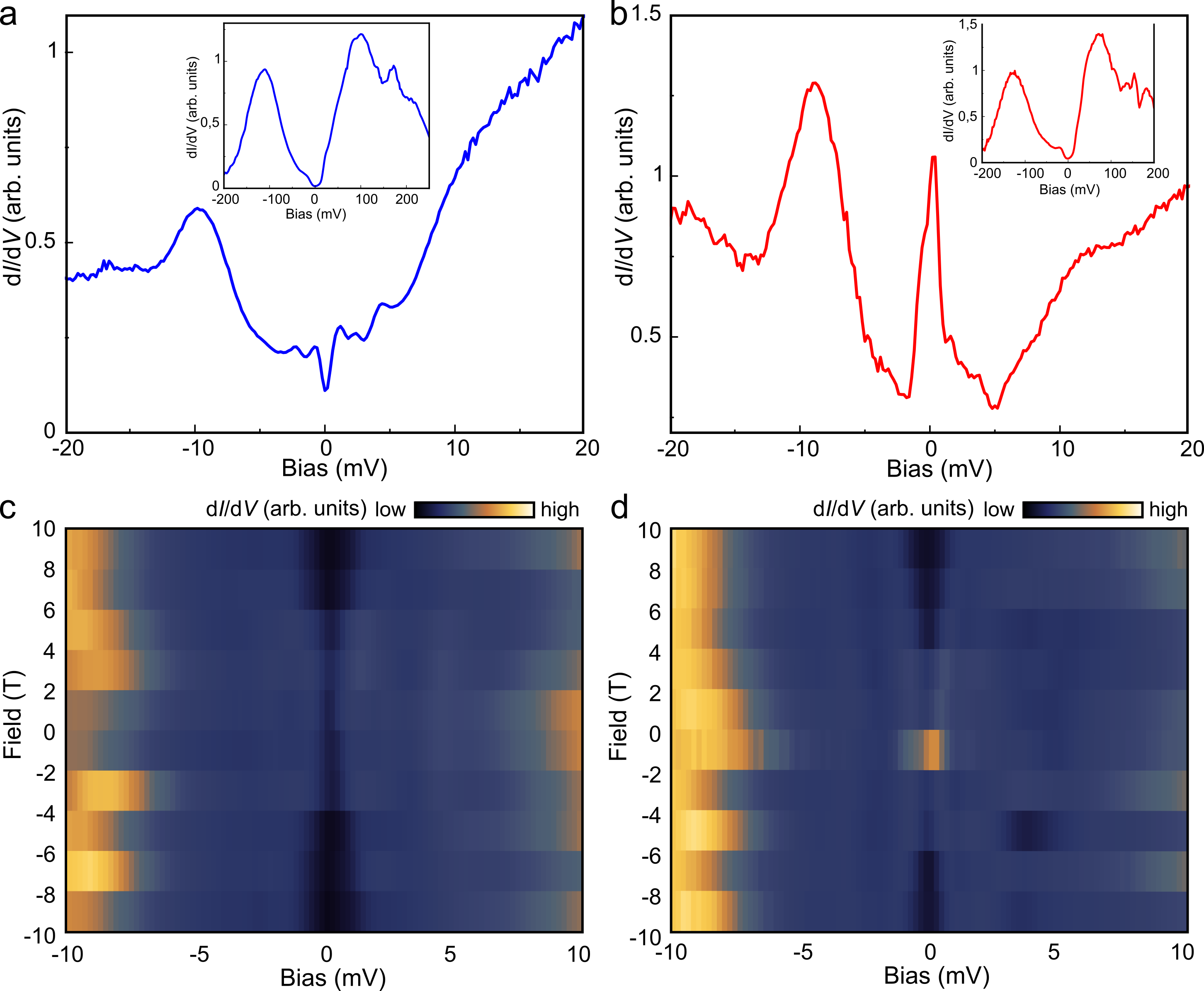}
    \caption{Magnetic field dependence of the half-filled sites with and without Kondo peak. Panel (a) shows half-filled site (large bias spectra shown in inset) without a Kondo peak. Panel (b) shows half-filled site (large bias spectra shown in inset) having Kondo peak. Panel (c, d) shows magnetic field dependence of sites shown in panel (a, b), respectively.}
    \label{fig:fig4}
\end{figure}

Having analyzed the spatial distribution of filled and half-filled sites, we now focus on the low energy spectroscopic response of the system. Half-filled sites host a local $S=1/2$. In the presence of antiferromagnetic coupling with the 1H substrate, this is expected to give rise to a Kondo peak. 1T/1H-TaS$_2$ and 1T/1H-TaSe$_2$ heterostructures realize half-filled Kondo lattices, where the coupling of the localized moments of 1T-layer and the conduction electrons in the 1H-layer result in heavy-fermion insulator or magnetically ordered states \cite{Vano2021,Wan2023,Ayani2023}.
In contrast, the 1T/1H-NbSe$_2$ heterostructure of our experiment has a large charge transfer leading to a strongly doped regime of the correlated lattice \cite{2023arXiv230214072C}, pushing our system towards the Kondo charge depletion found in 4Hb-TaS$_2$ \cite{Kumar_Nayak_2023,PhysRevLett.126.256402}. We observe that only $\sim6\%$ of half-filled sites have a pronounced Kondo peak (Fig.~\ref{fig:fig4}b), whereas most of the unit cells do not show clear Kondo signature (Fig.~\ref{fig:fig4}a). These different phenomenologies may stem from some half filled sites having antiferromagnetic Kondo coupling $J_K>0$, whereas others having ferromagnetic one $J_K<0$.
Local stacking is known to greatly impact the interlayer exchange coupling in van der Waals materials \cite{Chen2019,Soriano2019,Sivadas2018}, so that the different exchange couplings can naturally arise due to the different registry of 1T SoD and the $3\times3$ CDW of the underlying 1H-NbSe$_2$. The $3\times3$ CDW has been shown to strongly modulate the Yu-Shiba-Rusinov states of magnetic metal atoms on bulk 2H-NbSe$_2$ \cite{Liebhaber2019}, motivating this as a possible cause for our observations. Finally, it is unlikely that the absence of the Kondo features could be explained arising solely from impurities in 1H, given that the impurity density present in 1H is much lower that the one required to suppress Kondo in 94\% of the half filled sites.

In addition to the zero-magnetic field spectroscopy show in Fig.~\ref{fig:fig4}a, we have investigated the evolution of the spectroscopy with out-of-plane magnetic field of $\pm10$ T. The results are shown in Figs.~\ref{fig:fig4}c and d for the half-filled sites without and with Kondo features, respectively. Both sets of spectra feature a noticeable magnetic field dependence. Spectra corresponding to half-filled site with Kondo signature splits with magnetic field as expected. The low bias signature in half-filled site without Kondo also have a magnetic field dependent splitting. This response is very similar to the behaviour found on fully-filled sites (where Kondo related phenomena are not expected to occur, see SI) and could indicate that it would be associated with pseudogap enhancement of the underlying 1H-NbSe$_2$ substrate with applied field \cite{Ganguli2022,Vano2023}.

To summarize, we have shown that monolayer 1T-NbSe$_2$ develops a doped Mott insulating state due to charge transfer from the substrate. We have identified the appearance of both half-filled and fully-filled sites, showing that this material
provides a paradigmatic platform to emulate doped Hubbard physics in a triangular lattice.
Based on the statistical distribution of half- and fully-filled sites, we have found
signatures of underlying correlations, signaling that non-local correlations contribute to the
charge distribution of the doped Mott insulating monolayer.
From the perspective of strongly correlated phase diagram,
our experiment is likely to be in the overdoped regime, and
the emergence of superconducting state at lower dopings
would represent an open question.
Our results establish a new platform
to emulate doped Hubbard physics in a monolayer van der Waals material, 
providing a starting point to explore correlated phases of matter in geometrically
frustrated, tunable quantum materials.

\textbf{Acknowledgments}: This research made use of the Aalto Nanomicroscopy Center (Aalto NMC) facilities and was supported by the European Research Council (ERC-2017-AdG no.~788185 ``Artificial Designer Materials'') and Academy of Finland (Academy professor funding nos.~318995 and 320555, Academy research fellow nos.~331342, and 358088). We acknowledge the computational resources provided by the Aalto Science-IT project. X.H.~acknowledges the support from Huang Ruojun and Xiong Dongyan.

\bibliography{bib}

\end{document}